\documentclass[aps,prl,twocolumn,groupedaddress,showpacs]{revtex4}

\usepackage{graphicx}
\usepackage{amsmath}
\begin{document}


\title{Large-scale surface reconstruction energetics of Pt(100) and
  Au(100) by all-electron DFT} 


\author{Paula Havu$^1$}\thanks{Present address: Helsinki University of
  Technology (TKK), Finland}
\author{Volker Blum$^{1}$}
\author{Ville Havu$^{1,2}$}
\author{Patrick Rinke$^{1}$}
\author{Matthias Scheffler$^{1}$}
\affiliation{1) Fritz-Haber-Institut, Berlin, Germany}
\affiliation{2) Department of Applied Physics, Aalto University, Helsinki, Finland}


\date{\today}

\begin{abstract}
The low-index surfaces of Au and Pt all tend to reconstruct, a fact
that is of key importance in many nanostructure, catalytic,
and electrochemical applications. Remarkably, some significant questions    
regarding their structural energies remain even today, in particular for
the large-scale quasihexagonal reconstructed (100) surfaces:
Rather dissimilar reconstruction energies for Au and Pt in available 
experiments, and experiment and theory do not match for Pt. We here
show by all-electron density-functional theory that only large enough ``($5
\times$N)'' approximant supercells capture the qualitative
reconstruction energy trend between Au(100) and Pt(100), in contrast
to what is often done in the theoretical literature. Their magnitudes
are then in fact similar, and closer to the measured value for
Pt(100); our calculations achieve excellent agreement with known
geometric characteristics and provide direct evidence for the
electronic reconstruction driving force.
\end{abstract}

%
%

\maketitle

The late 5$d$ transition metals Pt and Au and their surfaces are of
central importance for a wide variety of applications, including
catalysis, electrochemistry, substrates for nanostructure creation and
characterization, etc. (some recent examples are discussed as Refs. \cite{Electrochemistry2006,Bond2006,Lucas09,Labayen03,Krug06,Baber10,
Lin10,Corso10,Casari09,Hernan98,Wilgocka10,Mateo-Marti10,Gao08,Suto06,Imbihl09,Lafferentz09,Bombis09,Alemani08,Dri08,Zandbergen07,Pierce09},
below). Their low-index surfaces (100),
(110) and (111) are paradigm systems to understand catalytic and
electrochemical processes at the atomic scale, to the point that only
textbooks and handbooks (e.g.,
Refs. \cite{Electrochemistry2006,Bond2006}) attempt somewhat 
comprehensive overviews. Au(111), (100) and Pt(100)
are ubiquitous as substrates for nanostructure creation and
manipulation, both due to their relative inertness and due to the fact that
they tend to form well-defined, large-scale \emph{quasihexagonal}
surface
\emph{reconstructions}\cite{Hagstrom65,Fedak66,Somorjai67,Bonzel72,Perdereau74,Sandy92}---not
just \emph{in vacuo} but also in solutions, under electrochemical
adsorption\cite{Lucas09} and epitaxial growth
conditions.\cite{Labayen03,Krug06} The reconstructions imply a
significant rearrangement of surface atoms, and a creation of
qualitatively distinct surface areas \cite{Binnig84,Ritz97} [fcc / hcp
stacking for (111), steeper vs. flatter ridges for (100)]. 
These local structure variations of Au and Pt (100) and (111) 
surfaces are the key ingredient for subsequent nanostructure growth
and processes on top, 
e.g., by way of a preferential nucleation of deposits (\cite{Baber10,Lin10,Corso10,Casari09,Hernan98} and
refs. therein). To give some recent  
examples where this atomic structure is important: the formation of magnetic 
structures\cite{Lin10,Wilgocka10}, self-assembled
nanostructured arrays of (bio)molecules,\cite{Mateo-Marti10}
magnetic dots\cite{Corso10} or supramolecules,\cite{Gao08,Suto06}
the structural and electronic underpinnings of catalytic
reactivity,\cite{Lucas09,Imbihl09,Baber10} on-surface synthesis
of well-separated molecular wires\cite{Lafferentz09} for transport
studies,\cite{Lafferentz09,Bombis09} well-defined patterns of
molecular switches,\cite{Alemani08,Dri08} or atomic-scale
visualization of kinetics and dynamics of surface
processes.\cite{Labayen03,Krug06,Zandbergen07,Pierce09} 
The balance of reconstruction/deconstruction can be
essential for chemical processes (\cite{Lucas09,Imbihl09} and
refs. therein); for instance, the peak of catalytic activity of
Pd/Au(100) occurs at a coverage of 0.07~ML,\cite{Chen05} precisely in
the range where one would expect reconstruction/deconstruction to occur.\cite{Hernan98,Blum99} 

In order to obtain a truly predictive theoretical understanding
of these phenomena, one would ideally like to use
current first-principles theory for ``realistic enough'' unit cell
sizes. However, this endeavour faces a significant 
obstacle: The sheer surface reconstruction size
necessitates that even very recent studies employ instead much smaller
model supercells.\cite{Mateo-Marti10,Gao08} This obstacle is most
pronounced for the (100) surfaces, whose quasihexagonal
superstructures are almost uniformly contracted compared to bulk (111)
planes [25\% more atoms per
  area in the ``hex'' layer than in a bulk-like (100) plane] and are,
strictly, incommensurate with the underlying square
substrate.\cite{Binnig84,Yamazaki88,Gibbs90,Gibbs91,Abernathy92,Abernathy93,Ritz97} 
\emph{In  vacuo}, they are additionally slightly rotated (0.5-1$^\circ$) against the
nominal (100) surface rows,\cite{Palmberg69,Heilmann79,Yamazaki88,Abernathy92,
  Abernathy93} but the exact ``hex'' layer geometry and surface
conditions are interrelated: On stepped surfaces, periodicities change,\cite{Moiseeva09}
and at high
$T$\cite{Heilmann79,Abernathy92,Abernathy93} or under catalytic 
conditions,\cite{Lucas09} unrotated ``hex'' planes are
observed. Again, this would not be a grave obstacle for 
either current experiments or theory \emph{if} suitably accurate,
small periodic approximants for the full surface structure could be
found. However, it turns out that the smallest approximants which
cover all key features (2-dimensional lateral
contraction, rotation, differently buckled qualitative surface areas)
are already significant in size: Roughly speaking, 
``(5$\times N$)'',\cite{Fedak66,Palmberg67,Palmberg69,Heilmann79} where
$N\!\!\approx\,$20-30, i.e., $\ge$100~atoms to be considered in each layer.

The primary point of the present paper is to demonstrate, from first
principles, that indeed large-scale ``(5$\times N$)'', and not smaller
approximants to the Au(100) and Pt(100) quasihexagonal reconstructions
are essential to obtain the qualitatively correct surface structural 
energetics from first principles. Our results are based on a
numerically converged all-electron description, using
density-functional theory (DFT) in the local-density (LDA) and PBE generalized
gradient approximation. While computationally challenging, such a
description is now feasible. In addition, it provides a key benchmark
for any computationally cheaper, more approximate methods to be used.

Concerning the reconstruction energy and driving force, some
significant \emph{qualitative} lessons can already be learned from 
simple (1$\times$1) surface models. Using DFT-LDA,
Takeuchi, Chan, and Ho (TCH) \cite{Takeuchi91} demonstrated that 
free-standing Au(111) planes achieve a large energy lowering when uniformly
contracted to the interatomic spacing of the Au(100) ``hex''
reconstruction. Using a Frenkel-Kontorova model, TCH found a net surface energy
gain through reconstruction, albeit very small:
$\Delta E_\text{hex}<$0.01~eV per unit area of the (100) surface
[eV/1$\times$1]. This is not far from an experimental, electrochemical estimate
(0.02~eV/1$\times$1 \cite{Santos04}), but, puzzlingly,
smaller by a factor $\approx$6 than a careful assessment of the reconstruction
energy of Pt(100) in vacuo, 0.12~eV/1$\times$1~\cite{Brown98} (an earlier
analysis placed Pt(100) even higher, at 0.21~eV/1$\times$1
\cite{Yeo95,Yeo96}). Adding to the
confusion is the fact that theoretical estimates for Pt(100) based on
small (5$\times$1) approximant cells \emph{in vacuo} yield only half
the experimental value (0.05-0.07~eV/1$\times$1
\cite{Chang99,Beurden04,Deskins05}). For Au(100), (5$\times$1)-based theoretical
estimates\cite{Feng05,Jacob07,Venkatachalam08} are 
approximately in line with the electrochemical experiments. Since
Au(100) and Pt(100) do behave similarly also 
regarding thermal stability,\cite{Abernathy92,Abernathy93} one might
consider the experimental reconstruction energy estimate for Pt(100)
an outlier---if it were not for the fact that it is \emph{this}
experiment which pertains to the full reconstruction in the
vacuum environment, and not to an electrochemical environment, or to a
restricted unit cell. Remarkably, our calculations of the full
(5$\times N$) reconstruction energy resolve this puzzle in favor of
the Pt(100) experiment, and show that both the experimental and
earlier theoretical values for
Au(100)\cite{Feng05,Jacob07,Venkatachalam08} underestimate the
reconstruction energy. 

\begin{figure}
  \includegraphics[width=0.40\textwidth]{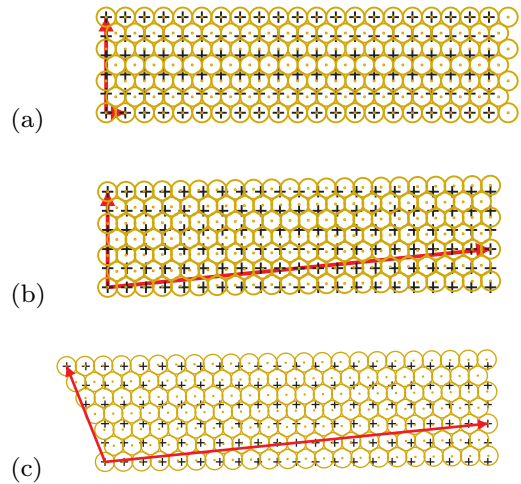}
  \caption{ \label{Fig:basic}
    Structure and distortion of the quasihexagonal layer. Circles:
    topmost (``hex'') layer, crosses: second (square) layer. Unit
    vectors of the coincidence lattice of both planes are shown in
    red.  (a) (5$\times$1)  and  (b) (5$\times$20)  
    superstructure. The ``hex'' layer is compressed by 4\% wrt. bulk
    (111) in both dimensions, and rotated by 1$^\circ$ in one
    dimension. (c) (5$\times$20)-like superstructure, rotated in both dimensions.
    }
\end{figure}

Figure \ref{Fig:basic} shows the unit cells of the specific
approximants that are important for this work. In each case,
the hexagonal top layer (circles) is slightly distorted to form a
commensurate coincidence lattice (arrows) with the underlying square
substrate (crosses). Subfigure (a) shows the popular
(5$\times$1) approximant. Compared to an ideal hexagonal layer, the
top layer is compressed by 4~\% in one dimension but not in the
other. To account for the reconstruction in both dimensions, larger
(5$\times N$)-type approximants ($N\approx$ 20-30) are needed. 
In subfigure (b), the hexagonal plane is thus compressed in
\emph{both} dimensions, and somewhat distorted so that its
close-packed rows are additionally rotated by $\approx 1^\circ$
against the substrate.
Finally, subfigure (c) shows a closely related (5$\times N$)-type
approximant, with a rotation of $\approx 1^\circ$ in both
dimensions. 
We model the ``hex'' layers as part of 
five-layer slabs (two layers relaxed, three kept fixed in bulk positions)
that are separated by more than 25~{\AA} of vacuum. The largest slabs ($N$=40)
thus comprise 1046 atoms, 446 of which are fully relaxed. We emphasize
that this number refers to heavy elements, no pseudoization is
employed, metallic systems are not amenable to current $O(N)$ methods, and that
especially transition metal slabs can be subject to serious
self-consistency instabilities (charge sloshing).\cite{Kresse96} On
the methodological side, our work therefore reflects 
significant efforts towards scalability and efficiency (including a
real-space version of Kerker preconditioning\cite{Blum09}) in our accurate ``DFT
and beyond'' code FHI-aims,\cite{Blum09,Havu09} which was used
throughout this work on massively parallel hardware (IBM's
BlueGene/P) and on a recent, infiniband-connected Sun Microsystems Linux
cluster. 

Compared to a (1$\times$1) layer, a (5$\times N$) plane ($N>$1) 
contains
($N$+5+1) additional atoms in the unit cell. In terms of total
energies for individual surface slabs, $E^\text{slab}$, and the total
energy per atom in the bulk, $E^\text{atom,bulk}$, 
the reconstruction energy $\Delta E_{5\times N}$ (here defined to be positive if a
reconstruction is favored) is
\begin{equation}\label{Eq:reconstruction}
  -\Delta E_{5\times N} = E^\text{slab}_{5\times N} - 5 N E^\text{slab}_{1\times 1}
                         - (N+6) E^\text{atom,bulk} \quad .
\end{equation}
For all energies in this work, we chose accurate computational
settings to guarantee a cumulative error of $\Delta E_{5\times N}$ below
0.02~eV/(1$\times$1) at most. Technical choices and convergence tests
are summarized in the EPAPS \cite{EPAPS} supporting material submitted
with this work. 

\begin{figure}
\includegraphics[width=0.4\textwidth]{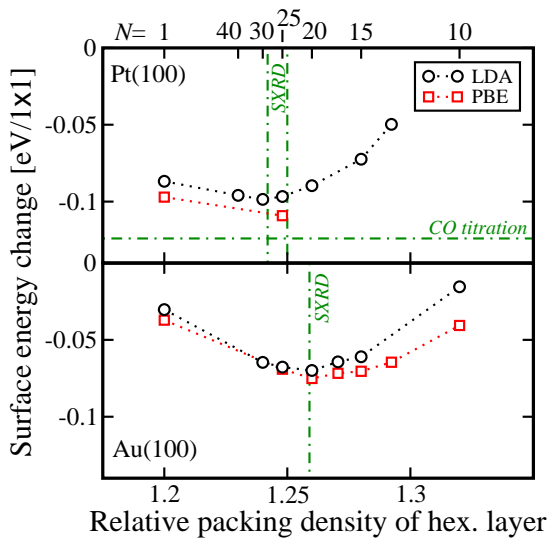}
  \caption{ \label{Fig:energetics}
    Change of surface energy with reconstruction, compared to
    (1$\times$1) [negative of $\Delta E_{5\times N}$,
    Eq. (\ref{Eq:reconstruction})], for Pt(100) (upper panel) and
    Au(100) (lower panel) using DFT-LDA/PBE, as a function of the
    lateral atomic packing density $\theta$. Dash-dotted
    lines: Experimental estimates for the packing density from
    SXRD,\cite{Abernathy92,Abernathy93} and for the Pt(100)
    reconstruction energy from the reanalysis \cite{Brown98} of CO
    titration calorimetry data.\cite{Yeo95,Yeo96} 
    }
\end{figure}
Figure \ref{Fig:energetics} summarizes our results regarding $\Delta
E_{5\times N}$ for Pt(100) and Au(100), using the approximant cells
Fig. \ref{Fig:basic}a and b. Since the density of (5$\times N$) planes
increases as $N$ decreases, we plot $\Delta
E_{5\times N}$ as a function of the \emph{lateral
atomic packing density} relative to the underlying (100) lattice:
$\theta=n_\text{hex}/n_{100}$. Experimentally, SXRD yields
$\theta\approx$1.242-1.250 for Pt(100)\cite{Abernathy92,Abernathy93} and 1.259
for Au(100).\cite{Gibbs90,Abernathy92} We find that the \emph{two-dimensional}
(5$\times N$) compression is indeed energetically favored over (5$\times$1) both in
LDA and PBE. In essence, the energy surface shows shallow
minima that are precisely in line with the experimental analysis. The
experimental trend of a denser ``hex'' layer for Au than for Pt is
clearly confirmed. For Pt, the energy gain of (5$\times N$) over (5$\times$1) is
small ($\sim$0.01~eV/1$\times$1), but Au roughly \emph{doubles} its 
reconstruction energy. 
This change brings the reconstruction energies of both surfaces close
to one another, and also to the 
experimental reconstruction energy estimate for Pt(100). In numbers,
we find $\Delta E_{5\times N}$=0.07 eV/1$\times$1 (Au)
vs. 0.10 eV/1$\times$1 (Pt) in DFT-LDA. In PBE, the reconstructed
surface energies are even slightly lower: for Pt, $\Delta E_{5\times
N}$=0.11 eV/1$\times$1 (PBE) is in remarkably close agreement with
experiment (0.12$\pm$0.02~eV/1$\times$1 \cite{Brown98}), and we recall
that we even expect a further lowering by $\approx$0.01~eV/1$\times$1
for the theoretical value from relaxation beyond the second layer, as
indicated by (5$\times$1) approximants with thicker slabs (see
EPAPS material\cite{EPAPS}). We thus conclude that \emph{there is no
  contradiction between
  theory and experiment for Pt(100)}, and that \emph{the
  reconstruction energies of Au(100) and Pt(100) are similar when the
  same in vacuo conditions are applied}. In our view, the seeming
agreement for Au(100)-(5$\times$1) theory \emph{in vacuo} and
electrochemical experiments was accidental. 

We next consider the energy difference between the
partially and fully rotated approximants in Figs. \ref{Fig:basic}b and
c for Au(100) in LDA at the optimum  packing density
(5$\times$20). The energy gain amounts to 4~meV/1$\times$1 in favor of
the approximant closest to experiment, Fig. \ref{Fig:basic}c. Between
a completely unrotated (5$\times$20) cell and the uniformly rotated
cell of Fig. \ref{Fig:basic}c, the change would be 8~meV/1$\times$1.
This number is still significant when taken per
reconstructed unit cell, but is also an order of magnitude smaller than
the full reconstruction energy. It is thus fully consistent with the
observation that the hex plane can be \emph{modified} depending on its environment,
\cite{Lucas09,Moiseeva09} without disrupting the reconstruction
\emph{per se}, and gives a quantitative idea of the energy scale of
these processes.

\begin{figure}
  \includegraphics[width=0.42\textwidth]{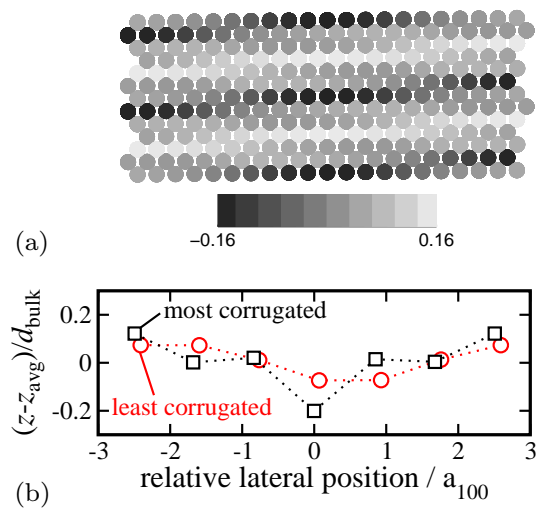}
  \caption{\label{Fig:top-view}
           (a) Top view of the ``hex'' layer on an
           Au(100)-''(5$\times$20)'' reconstruction model
           (LDA). $z$ coordinates of all atoms are color-coded in
           units of $d_\text{bulk}$, relative to the average $z$ position
           of the plane. (b) Height
           profiles through the most corrugated and least corrugated
           surface areas. 
          }
\end{figure}
In the accompanying EPAPS material,\cite{EPAPS} we demonstrate the
excellent agreement of the overall reconstruction geometry calculated
here with experimental characteristics available in the literature. To
illustrate the consequences for local surface structure variations,
Fig. \ref{Fig:top-view} shows a top view
(``hex'' layer only) of the 
``(5$\times$20)'' reconstructed Au(100) surface in LDA. Qualitatively,
this picture looks strikingly similar to published atomically resolved
STM images.\cite{Ritz97} For a more
quantitative view, the frame below displays the $z$ coordinate of
atomic zigzag rows in the ``5'' direction with the largest and smallest
corrugation amplitude, respectively. The variation 
shows that very different local environments exist in the 
surface as a result of the \emph{two-dimensional} reconstruction. The
maximum corrugation in the ``5'' direction is 0.65~{\AA} (32\% of 
the bulk interlayer spacing $d_\text{bulk}$), compared to only 15\% in
the least corrugated area. These numbers are 
very similar in LDA and PBE, and also very similar to Pt(100)-(5$\times$25)
(29\% and 14\%, respectively). It is precisely such
different local environments that drive the preferential nucleation
and nanostructuring phenomena on such surfaces.\cite{Baber10,Lin10,Corso10,Casari09,Hernan98}

Finally, we turn briefly to the electronic-structure changes that
accompany the reconstruction, which we can address directly; such
changes are the basis to understand the role of
catalytic activity in 5$d$ based systems.\cite{Baber10} For 5$d$ 
metals, relativity enhances the 
participation of the valence $d$ electrons by shifting them upwards
towards the $s$ levels.\cite{Desclaux76,Pyykko04} Since $d$
electrons are thus more easily promoted to $sp$-like states, enhanced
$sp$ bonding has been proposed as a cause of the reconstruction.\cite{Rhodin69,Annett89} 
On the other hand, enhanced bonding among the $d$ states themselves was
favored in Ref. \cite{Takeuchi91}.
\begin{table}
  \begin{tabular}{cccccc}
    \hline\hline
      Material          &   Ir       & Pt      & Au      & Au(NR) & Ag \\
    \hline
      Reconstruction    & 5$\times$1 & ``hex'' & ``hex'' & none   & none \\
      Band center shift [eV] & $-$0.19    & $-$0.24 & $-$0.13 & $-$0.01& $-$0.08 \\
    \hline
  \end{tabular}
  \caption{\label{Tab:DOS}
     Valence band center shift of the projected densities of
     states (LDA) of the surface atoms between (100)-(1$\times$1)
     and (5$\times$1) ``hex'' approximants, for various
     elements, compared to the computed reconstruction
     tendency. Au(NR) denotes Au, but with a non-relativistic kinetic
     energy.  
  }
\end{table}
Table \ref{Tab:DOS} shows the shift of the valence band centers of
the surface layer projected densities of states of Ir(100), Pt(100),
and Au(100) (all three reconstruct) and Au(100) without relativity, or
Ag(100) (the latter two surfaces would not reconstruct), when going
from a (1$\times$1) to a (5$\times$1) reconstruction. In short, a
characteristic downward shift of the valence band center occurs for
surfaces that reconstruct, but is much smaller for the other two
surfaces. An additional angular momentum decomposition of the
changes shows only a tiny transfer of electrons between the $d$ and
$sp$ channels, so the associated change must happen within the $d$ states
themselves. Taken together, this analysis thus supports the
relativistically enhanced $d$-$d$ hybridization mechanism suggested
by TCH \cite{Takeuchi91} over $sp$ promotion.\cite{Rhodin69,Annett89} 

In conclusion, we have demonstrated that large reconstruction
approximants are indeed necessary to capture the subtle reconstruction
energy balance of Au(100)- and Pt(100)-"hex" quantitatively. Together
with the increasing power of computers and computational methods,
realistically large simulations of chemical processes in the presence
of reconstruction/deconstruction are now within reach.


\end{document}